\documentstyle[12pt]{article}
\begin{document}
% definitions
\newcommand{\newc}{\newcommand}

\newc{\be}{\begin{equation}}
\newc{\ee}{\end{equation}}
\newc{\ba}{\begin{eqnarray}}
\newc{\ea}{\end{eqnarray}}
\newc{\ie}{{\it i.e.} }
\newc{\eg}{{\it e.g.} }
\newc{\etc}{{\it etc.} }
\newc{\etal}{{\it et al.} }
\newc{\ra}{\rightarrow}
\newc{\lra}{\leftrightarrow}
\newc{\no}{Nielsen-Olesen }
\newc{\lsim}{\buildrel{<}\over{\sim}}
\newc{\gsim}{\buildrel{>}\over{\sim}}
\begin{titlepage}
\begin{center}

\vskip 4.5cm

{\large \bf
Existence and Stability of Spinning Embedded Vortices
}
\vskip .5cm
{\it Talk presented at the NATO Advanced Research Workshop `Electroweak
Physics and the Early Universe',Sintra, Portugal, March 23-25, 1994.}
\vskip .4in
{\large Leandros Perivolaropoulos}\footnote{E-mail address:
leandros@cfata3.harvard.edu},\footnote{Also, Visiting Scientist,
Department of Physics, Brown University, Providence, R.I. 02912.}\\[.15in]

{\em Division of Theoretical Astrophysics\\
Harvard-Smithsonian Center for Astrophysics\\
60 Garden St.\\
Cambridge, Mass. 02138, USA.}
\end{center}
\vspace{1cm}
\begin{abstract}
I show the existence of a new type of vortex solution which is non-static but
stationary and carries angular momentum. This {\it spinning vortex} can be
embedded
in models with trivial vacuum topology like a model with $SU(2)_{global}\times
U(1)_{local}
\rightarrow U(1)_{global}$ symmetry breaking. The stability properties of the
embedded
spinning vortex are also studied in detail and it is shown that stability
improves
drastically as angular momentum increases. The implications of this result for
vortices
embedded in the electroweak model are under study.
\end{abstract}
\end{titlepage}

This work focuses on a new class of embedded vortices which are non-static
but stationary and carry non-zero angular momentum. They may
therefore be called {\it spinning embedded vortices}. It will be shown that
these vortices have improved stability properties compared to their
non-spinning
counterparts (electroweak vortices with in a hypothetical model with
Weinberg angle $\theta_w = \pi
/2$\cite{va91,v92,jpv92}) whose properties are discussed in the contributions
of
Klinkhamer and Vachaspati in this volume.

Let's first clarify the motivation for generalizing the concept of the
non-spinning electroweak (hereafter EW) vortex. As is well known the vacuum
manifold of the standard EW model is a three sphere $S^3$. This implies that
the model does not support {\it any} stable topological defects. However,
almost a couple of years ago, Vachaspati\cite{v92} pointed out that there is a
vortex-like coherent state in the EW model which appears as an exact
solution to the field equations. In a later paper\cite{jpv92} we showed that
this coherent state, called the embedded electroweak vortex\cite{vb92}, is
in fact stable for a finite sector in parameter space. We also showed that
the stability sector does not include the physically realized values of
parameters. This was unfortunate but raised an exciting new question:
{\it Are there different types of embedded vortices with improved stability
properties?}.

Until recently there had been two main classes of attempts to address this
question. The first class originated from a work of Vachaspati and
Watkins\cite{vw93} who showed that bound states on the embedded vortex may
indeed improve its stability. Due to the complexity of the problem however,
it was not clear if this improvement of stability was enough to stabilize
the electroweak vortex for the physical values of parameters. The second
class focused on embeddings in {\it extensions} of the standard EW model.
There have been several interesting works on this
subject\cite{l93,p93,ds93,ej93} but of
particular interest is the work of Dvali and Senjanovic\cite{ds93} who
discovered {\it topologically stable} vortices in the two Higgs doublet EW
model.

Here I focus on a new stabilization mechanism which
is applicable to vortices of the standard, minimal EW model and
introduces {\it spin}
to improve the stability of the embedded vortices. An efficient way to
introduce angular momentum to the vortex configuration is to embed it in a
background of charge density which may be provided for example, by a charged
external field with coherence length much larger than the width of the
vortex.

Consider a toy model with symmetry breaking $SU(2)_{global} \times U(1)_{local}
\longrightarrow U(1)_{global}$.
For our purpose, this model is identical to the bosonic sector of the
standard EW model with $\theta_w =\pi/2$.
The vacuum manifold in this model is $S^3$.
It has been shown\cite{va91,h92} that the \no vortex\cite{no73} is supported as
an exact
solution and that it is stable for a limited parameter range.
Consider now an embedding ansatz\cite{p94} which is a stationary
generalization\cite{d90}
of the \no ansatz
\be
\Phi=\left( \begin{array}{c}
0\\
f(r) e^{i m \theta} e^{i\omega_0 t}
\end{array} \right)
\ee
\be
A_\theta = {{v(r)}\over r}, \hspace{1cm} A_0 = \alpha (r)
\ee
The differences between the ansatz (1), (2) and the embedded \no
vortex\cite{v92} (EW string with $\theta_w = \pi /2$) is the linear
time dependence of the Goldstone field and the allowed possibility of having
$A_0 \neq 0$. The Lagrangian density which supports this ansatz as a
solution is a generalized Abelian Higgs Lagrangian where the scalar field
$\Phi$ has been promoted to an $SU(2)$ doublet and also a background charge
density has been introduced and coupled to the gauge field $A_\mu$.
It is of the form
\be
{\cal L}=-{1\over 4} F_{\mu \nu}F^{\mu \nu} +{1\over 2} \vert D_\mu \Phi \vert
^2
-{\lambda \over 4} (\vert \Phi \vert ^2 - \eta ^2)^2 - A_\mu J^\mu
\ee
where  $J^\mu = (\rho (r),0,0,0)$ is a background charge density.
 Let $\rho (r) \rightarrow \rho_0$ asymptotically, where $\rho_0$ is a
constant. The two crucial parameters that determine the stability of the
embedded vortex (1), (2) will be shown to be $\beta$ and $\rho_0$. The
angular velocity $\omega_0$ is fixed in terms of those parameters. The
total charge of the vortex configuration plus the background is
\be
Q=\int_V d^2 x\hspace{1mm} [f^2 \hspace{1mm}
(\omega_0-\alpha)  +
\hspace{1mm} \rho (r) ]
\ee
where
the first term in the integral is due to the vortex field and the second is due
to the
background. This charge is in general non-zero, finite and conserved. The
angular momentum ${\vec M_v}$ of the vortex configuration may be shown to be
proportional to the charge $Q_v$ and the winding number $m$ of the vortex
field
\be
{\vec M_v}=\int_V d^2 x \hspace{1mm}{\vec r} \times ({\vec E_v}\times {\vec
B})=-2\hspace{1mm} \int_V
d^2 x
\hspace{1mm}
f(r)^2 \hspace{1mm} (\omega_0-\alpha) \hspace{1mm} v(r) \hspace{1mm} {\hat e}_z
\simeq -2
\hspace{1mm}  m\hspace{1mm}Q_v \hspace{1mm} {\hat e}_z
\ee

The crucial question that needs to be addressed is, {\it what are the
stability properties of the vortex and how do they compare with the
corresponding properties of its non-spinning counterpart (embedded EW vortex
with $\theta_w =\pi /2$)}?

Consider a perturbation to the ansatz (1), (2) of the form
\be
\delta \Phi  =
\left(
\begin{array}{c}
  g(r) e^{i n \theta} \\
\delta f (r, \theta )
\end{array}
\right)
, \hspace{1cm} \delta A_\mu (r,\theta)
\ee
The energy of the perturbed configuration decouples as follows
\be
E= E_0 (f,v,\alpha) + \delta E (\delta f, \delta A_\mu) + E_1 (g)
\ee
where the energy $E_0$ of the unperturbed configuration is identical in
form to the energy of the topologically stable configuration obtained when
the complex doublet in the ansatz is replaced by a complex singlet. The term
$\delta E$ is identical in form to the perturbation of the topologically
stable spinning vortex and must be positive definite since the topological
vortex is stable. Thus the term whose sign determines the stability is the
term $E_1 (g)$. For those parameter values $(\beta, \rho_0)$ for which $E_1
(g)$ can be negative, the embedded vortex (1), (2) is unstable. But $E_1$ may
be shown\cite{p94} to be
positive definite {\it iff} the eigenvalue equation
\be
-g''- {{g'}\over r} + {(v-n)^2 \over r^2} g + \beta (f^2-1)g  + \alpha^2 g=
\omega^2 g
\ee
has no negative eigenvalues. For the range of parameters $(\beta, \rho_0)$
for which this happens, the embedded spinning vortex is stable.

The unperturbed fields, $f,v,\alpha$ and therefore the sign of $E_1$, also
depend on the detailed form of $\rho (r)$ close to the core of the vortex. I
will therefore consider two opposite extreme forms of $\rho (r)$ and show
that stability is dramatically improved in both cases. For the first type of
background, the {\it soft} background, $\rho (r)$ adjusts in a way to
completely neutralize the charge induced by the spin of the vortex \ie
\be
\rho (r)=-\omega_0 f^2 (r), \hspace{1cm} \alpha=0
\ee
 For such a background, the field
equations allow $\alpha=0$ everywhere. This implies that the
Schroedinger-type potential of the eigenvalue equation (8), looses a positive
definite contribution: the term $\alpha^2$. This makes the equation more
receptive to negative eigenvalues. We are therefore led to the conjecture
that the soft background is the background of minimum stability.

The other opposite extreme form is that of a {\it hard} background where
$\rho (r)$ does not feel the presence of the vortex and is constant
everywhere
\be
\rho (r) = \rho_0, \hspace{1cm} \alpha \neq 0
\ee
 In that case, $A_0$ (or $\alpha$) can not vanish\cite{p94} and we expect
better
stability properties for this background type.

\vspace{6cm}

{\bf Figure 1:} The stability map is parameter space for the embedded
spinning vortex.
Sector III. (II. and III.) is the stability sector for the
case of a soft (hard) background.

\vspace{0.5cm}

It is straightforward to solve the eigenvalue problem numerically
using relaxation methods\cite{nr}, for the two types of
backgrounds and map the stability sectors in parameter space $(\beta,
\rho_0)$ for each case. The stability sector for the soft (hard) background
is on the right of the continous (dotted) line in Fig. 1.

The case of the
non-spinning EW vortex is obtained for $\rho_0 =0$ and has been studied
previously \cite{h92,jpv92} (stability only for $\beta < 1$).
Clearly the range of $\beta$ corresponding to
stability increases dramatically when spin is introduced ($\rho_0 \neq 0$).
In fact it may be shown analytically\cite{p94} that for $\rho_0 \rightarrow
\infty$ the embedded spinning vortex is stable for {\it any} $\beta$.

In conclusion, I have shown that angular momentum induced by a charged
background can indeed stabilize the embedded EW vortex. This result raises
some interesting new questions. First, {\it can spin stabilize the EW vortex
for physical values of parameters?}. Second, {\it can global embedded vortices
(which are normally always unstable\cite{bpv}) be stabilized by the
introduction of
spin?}
 These issues are currently under investigation.


\vskip 1cm

\begin{thebibliography}{99}
\bibitem{va91} T. Vachaspati and A. Achucarro, {\it Phys. Rev.} {\bf
D44}, 3067 (1991).
\bibitem{v92}  T. Vachaspati, {\it Phys. Rev. Lett.} {\bf 68}, 1977 (1992).\\
T. Vachaspati, {\it Nucl. Phys.} {\bf B397}, 648
(1993).
\bibitem{jpv92} M. James, L. Perivolaropoulos and T. Vachaspati,
{\it Phys. Rev.}{\bf D46}, 5232 (1992)\\
M. James, L. Perivolaropoulos and T. Vachaspati,
{\it Nucl. Phys.} {\bf B395}, 534 (1993).
\bibitem{vb92} T. Vachaspati and M. Barriola, {\it Phys. Rev. Lett.} {\bf 69},
1867 (1992).
\bibitem{vw93} T. Vachaspati and R. Watkins, {\it Phys. Lett.} {\bf B318}, 163
(1993).
\bibitem{l93} H. S. La, hep-ph/9302220 (1993).
\bibitem{p93} L. Perivolaropoulos, {\it Phys. Lett.} {\bf B316}, 328 (1993).
\bibitem{ds93} G. Dvali and G. Senjanovic, {\it Phys. Rev. Lett.} {\bf 71},
2376 (1993).
\bibitem{ej93} M. Earnshaw and M. James, {\it Phys. Rev.} {\bf D48}, 5818
(1993).
\bibitem{no73} H. B. Nielsen and P. Olesen, {\it Nucl. Phys.} {\bf B61}, 45
(1973). \\
see also, L. Perivolaropoulos, {\it Phys. Rev.} {\bf D48}, 5961 (1993)
for a correction in the vortex asymptotic behavior presented by
Nielsen-Olesen and in most subsequent papers.
\bibitem{p94} L. Perivolaropoulos, {\it Phys. Rev.} {\bf D} in press.
\bibitem{h92} M. Hindmarsh, Phys. Rev. Lett. {\bf 68}, 1263 (1992)\\
A. Ach\'ucarro, K. Kuijken, L. Perivolaropoulos
T. Vachaspati, {\it Nucl. Phys.} {\bf B388}, 435 (1992).
\bibitem{d90} R. Davis, {\it Mod. Phys. Lett.} {\bf A5}, 955 (1990).
\bibitem{nr} W. Press, B. Flannery, S. Teukolsky, W. Vetterling,
{\it Numerical Recipes, 2nd Edition},  Cambridge Univ. Press (1993).
\bibitem{bpv} M. Barriola, L. Perivolaropoulos and T. Vachaspati, unpublished.
\end{thebibliography}
\end{document}